\documentclass[aps,prl,twocolumn,superscriptaddress,reprint]{revtex4-1}
\usepackage{graphicx}
\usepackage{setspace}
\usepackage{amsmath}
\usepackage{amssymb}
\usepackage{color}
\usepackage{array}
\usepackage{subfigure}
\usepackage{hyperref}
\usepackage{float}
\usepackage{lipsum}

\usepackage[all]{xy}
\newcommand{\RN}[1]{%
  \textup{\uppercase\expandafter{\romannumeral#1}}%
}

\begin{document}
%\begin{CJK*}{GB}{}
\title{Polarization induced interference within electromagnetically induced transparency for atoms of double-V linkage}
\author{Yuan Sun}
\email[email: ]{sunyuan17@nudt.edu.cn}
\affiliation{Interdisciplinary Center for Quantum Information, National University of Defense Technology, Changsha 410073, P.R.China}
\author{Chang Liu}
\affiliation{Institute for Quantum Computing and Department of Physics \& Astronomy, University of Waterloo, Waterloo, Ontario N2L3G1, Canada}
\author{Ping-Xing Chen}
\affiliation{Interdisciplinary Center for Quantum Information, National University of Defense Technology, Changsha 410073, P.R.China}
\author{Liang Liu}
\email[email: ]{liang.liu@siom.ac.cn}
\affiliation{Key Laboratory of Quantum Optics and Center of Cold Atom Physics, Shanghai Institute of Optics and Fine Mechanics, Chinese Academy of Sciences, Shanghai 201800, P.R.China}

\begin{abstract}
People have been paying attention to the role of atoms' complex internal level structures in the research of electromagnetically induced transparency (EIT) for a long time, where the various degenerate Zeeman levels usually generate complex linkage patterns for the atomic transitions. It turns out, with special choices of the atomic states and the atomic transitions' linkage structure, clear signatures of quantum interference induced by the probe and coupling light's polarizations can emerge from a typical EIT phenomena. We propose to study a four state system with double-V linkage pattern for the transitions and analyze the polarization induced interference under the EIT condition. We show that such interference arises naturally under mild conditions on the optical field and atom manipulations. Moreover, we construct a variation form of double-M linkage pattern where the polarization induced interference enables polarization-dependent cross-modulation between incident lights that can be effective even at the few-photon level. The theme is to gain more insight into the essential question: how can we build non-trivial optical medium where incident lights will induce polarization-dependent non-linear optical interactions while covering a wide range of the incidence intensity from the many-photon level to the few-photon level, respectively.
\end{abstract}
\maketitle
%\end{CJK*}

Ever since the early days of investigating EIT and four-wave mixing (FWM) in atomic mediums, it has been noticed that the effects of the multiple Zeeman sub-levels can not be overlooked \cite{PhysRevA.53.1014, PhysRevA.61.053805, PhysRevA.63.063406, PhysRevA.81.033801, LaserPhys.24.094011, PhysRevA.94.043822}. Under special scenarios such a multi-state system with complicated linkage structure can be transformed back into a simple three level EIT system \cite{PhysRevA.27.906, PhysRevA.74.053402, PhysRevA.76.033817}, while typically it leads to complexities in the EIT profile or FWM process which usually exhibit strong correlations with the polarizations of the incident light \cite{PhysRevA.74.033821, PhysRevA.77.051803, PhysRevA.81.033801, PhysRevA.85.051805, Zheng:09_MinXiao_OE, LaserPhys.24.094011}. Investigations into these effects have already leaded to a few interesting findings, such as the controlled rotation of the polarization of an incident optical pulse \cite{PhysRevA.74.033821}, manipulation of the transparency window \cite{PhysRevA.81.023830, PhysRevA.85.051805, LaserPhys.24.094011},  vector magnetometry from EIT in linearly polarized light \cite{PhysRevA.82.033807}, and controlling enhancement or suppression of FWM by polarized light \cite{PhysRevA.81.033801}. Of particular importance is the construction of quantum memory for photon polarization states via utilizing those Zeeman sub-levels \cite{PhysRevLett.103.043601, PhysRevLett.111.240503}. Those previous investigations have paved the way for studying the quantum interference induced by the polarizations of the driving lasers with special linkage geometry of atomic states in optically thick medium formed by cold alkali atoms.
\par
Meanwhile, generating and manipulating nonlinear interactions between optical fields of low intensities at the few-photon level is of essential importance in the research frontier of quantum optics \cite{PhysRevLett.84.1419}. According to early predictions of Harris and Hau, large cross-phase shifts through enhancing the weak Kerr effect by EIT at very low light intensities down to the single-photon level are hardly attainable \cite{PhysRevLett.82.4611}. In particular, a lot of efforts have been devoted to enhancement of the nonlinear interaction between single photon pulses via high-finesse cavity \cite{PhysRevLett.75.4710, VIT_science1266, RevModPhys.87.1379} or Rydberg blockade \cite{PhysRevLett.107.133602, PhysRevA.94.053830, RevModPhys.82.2313}. Meanwhile, people are trying to seek novel schemes based upon EIT and FWM for cross-modulation at the single-photon level without revoking the experimental complexities of cavities or Rydberg atoms. Many exciting progresses have been achieved along this direction, including the double slow light method in multi-level system \cite{PhysRevLett.101.073602, PhysRevLett.106.193006}, stationary light method \cite{PhysRevLett.108.173603}, and especially the recent developments of cross-phase modulation (XPM) via FWM of double-$\Lambda$ configuration in a three level system \cite{PhysRevLett.115.113005} or four level system \cite{PhysRevLett.117.203601}. Although the photonic polarization degree of freedom has not been explicitly brought into these XPM schemes so far, they have naturally triggered the motivation of inducing polarization-dependent nonlinear interactions between optical fields of low intensities down to the few-photon level via EIT and FWM methods.
\par
In this letter, we discuss the polarization induced quantum interference under the general EIT condition in a special double-V linkage structure of atomic internal electronic states, which can be realized in $^{87}$Rb atom. We demonstrate in theory that this interference is inherently associated with the polarization degree of freedom, has clear physical signature and can be observed with moderate experimental conditions, which has potential applications in all-optical polarization control. Then we extend this concept to a multi-state system with double-M linkage structure, where we construct a mechanism of polarization-dependent cross-modulation between the two incident lights. The proposed mechanism is in principle applicable to very low incident intensities such as a weak coherent light pulse containing energy equivalent to only a few photons. Throughout this letter, all problems are treated with the one-dimensional approximation \cite{RevModPhys.82.1041}, where effectively only on transverse mode is taken into consideration.
\par
\begin{figure}[t]
 \centering
\begin{tabular}{l}
\includegraphics[trim = 0mm 0mm 0mm 0mm, clip, width=8.5cm]{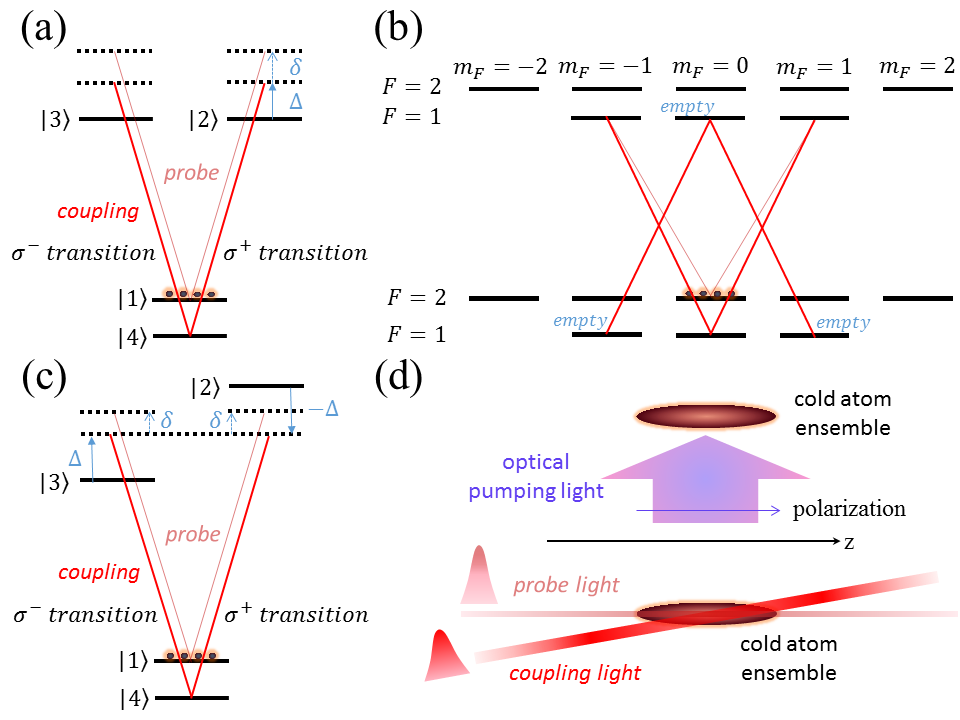}
%left down right up
\end{tabular}
\linespread{1} 
\caption{Polarization-resolved configuration of EIT in atoms with double-V linkage pattern. (a) Schematic of the double-V linkage structure of a simplified four-state system where the two excited states are degenerate in energy. The $k$-vectors of incident coupling and probe lights are along the z-direction such that they couple to the $\sigma^+$-transitions and $\sigma^-$-transitions respectively. The probe beam is assumed to have relatively low intensity hence it does not disturb the population of state $|1\rangle$. (b) Implementation of the abstracted double-V linkage structure in $^{87}$Rb D1 or D2 transitions, where $|1\rangle$ and $|4\rangle$ are realized by the two magnetic insensitive clock states. It can also be implemented in Rydberg EIT system \cite{PhysRevA.94.043822, SuppInfo}.
(c) A double-V linkage structure where $|2\rangle$ and $|3\rangle$ are not degenerate in energy, for example, when an external magnetic field along z-direction induces first order Zeeman shift. (d) Simplified proposal for experimental implementation. An optical pumping light shines onto the optically dense cold atom ensemble to begin with, which prepares the initial state. Probe beam and coupling beam are assumed to be parallel and overlapping, or with a tiny angle for the purpose of phase matching in the medium if desired. \label{schematic_1}}
\end{figure}

More specifically, consider an EIT process where the polarizations of the probe and coupling lights are resolved such as Fig. \ref{schematic_1}(a)\&(b), where an optical pumping process concentrates the initial population into state $|1\rangle$. The state initialization is necessary to remove the requirement of $J=0$ of Ref. \cite{LaserPhys.24.094011} which severely limits the choice of atoms to special ones like thallium. When the probe light intensity is weak, its equation of motion (EOM) up to the first order can be derived from the Maxwell-Bloch equations in the rotating wave frame as the following:
\begin{subequations}
\label{1a_EOM}
\begin{align}
\frac{\partial}{\partial z} \Omega_{p+} + \frac{1}{c} \frac{\partial}{\partial t} \Omega_{p+} = i \frac{n\sigma\Gamma}{2} \rho_{21};
\\
\frac{\partial}{\partial z} \Omega_{p-} + \frac{1}{c} \frac{\partial}{\partial t} \Omega_{p-} = i \frac{n\sigma\Gamma}{2} \rho_{31};
\\
\label{1a_EOM_c}
\frac{d}{dt} \rho_{21} = \frac{i}{2}\Omega_{p+} - \frac{i}{2}\frac{\Omega_c}{\sqrt{2}}\rho_{41} + (i(\Delta+\delta) - \frac{\Gamma}{2})\rho_{21};
\\
\frac{d}{dt} \rho_{31} = \frac{i}{2}\Omega_{p-} + \frac{i}{2}\frac{\beta_c\Omega_c}{\sqrt{2}}\rho_{41} + (i(\Delta+\delta) - \frac{\Gamma}{2})\rho_{31};
\\
\frac{d}{dt} \rho_{41} = -\frac{i}{2}\frac{\Omega^*_c}{\sqrt{2}}\rho_{21} 
+\frac{i}{2}\frac{\beta^*_c\Omega^*_c}{\sqrt{2}}\rho_{31}
+ (i\delta -\frac{\gamma}{2})\rho_{41};
\end{align}
\end{subequations}
where subscript p, c stands for probe and coupling lights respectively, $\Delta$ is the one-photon detuning, $\delta$ is the two photon detuning, $\Gamma$ is the decay rate of $|2\rangle$ and $|3\rangle$, $\gamma$ is the decoherence rate between $|1\rangle$ and $|4\rangle$, $n$ is the atom density, and $\sigma$ is the atom-light cross section for the probe light. $\Omega_{p+}, \frac{\Omega_c}{\sqrt{2}}$ are the Rabi frequencies for the $\sigma^+$ transitions while $\Omega_{p-}, \frac{\beta_c\Omega_c}{\sqrt{2}}$ are the Rabi frequencies for the $\sigma^-$ transitions, with $|\beta_c|=1$. Namely, the coupling light is assumed to be linearly polarized with decomposition via the circular polarization basis: $P_c = \frac{1}{\sqrt{2}}(P_+ + \beta_cP_-)$. This assumption of linear polarization is not necessary and the analysis below can be easily extended to any polarization state of the coupling light \cite{SuppInfo}.
\par
We examine the propagation of the probe beam through the cold atom medium of finite optical depth (OD) along the z-direction, by solving for the steady state solution of Eq.\eqref{1a_EOM} \cite{RevModPhys.77.633, SuppInfo}; then the equation governing the dynamics of the probe light's two polarization components is obtained as the following:
\begin{equation}
\label{1a_dynamics_1}
\frac{d}{dz}
\begin{bmatrix}
\Omega_{p+} \\
\Omega_{p-}
\end{bmatrix}
= \frac{i}{2} n \sigma \Gamma A^{-1}
\begin{bmatrix}
\Omega_{p+} \\
\Omega_{p-}
\end{bmatrix}, 
\end{equation}
where $A$ is a 2 by 2 matrix independent of $z$, with $A_{11}=A_{22}=\frac{\Omega^2_c}{2(2\delta+i\gamma)}-2\Delta-2\delta-i\Gamma$, $A_{12} = -\beta^*_c \frac{\Omega^2_c}{2(2\delta+i\gamma)}$ and $A_{21} = -\beta_c \frac{\Omega^2_c}{2(2\delta+i\gamma)}$.
\par
The matrix $A$ has eigenvalues $\lambda_1, \lambda_2$ as in the following equation; and therefore Eq.\eqref{1a_dynamics_1} can be decoupled into two independent branches which correspond to two types of different propagation dynamics of the probe light:
\begin{subequations}
\label{1a_eigenvalues}
\begin{eqnarray}
\label{1a_eigenvalues_a}
\lambda_1 = -2\Delta-2\delta-i\Gamma, \\
\label{1a_eigenvalues_b}
\lambda_2  = \frac{\Omega^2_c}{2(2\delta+i\gamma)}-2\Delta-2\delta-i\Gamma.
\end{eqnarray}
\end{subequations}
\par
\begin{figure}[t]
 \centering
\begin{tabular}{l}
\includegraphics[trim = 0mm 0mm 0mm 0mm, clip, width=7.5cm]{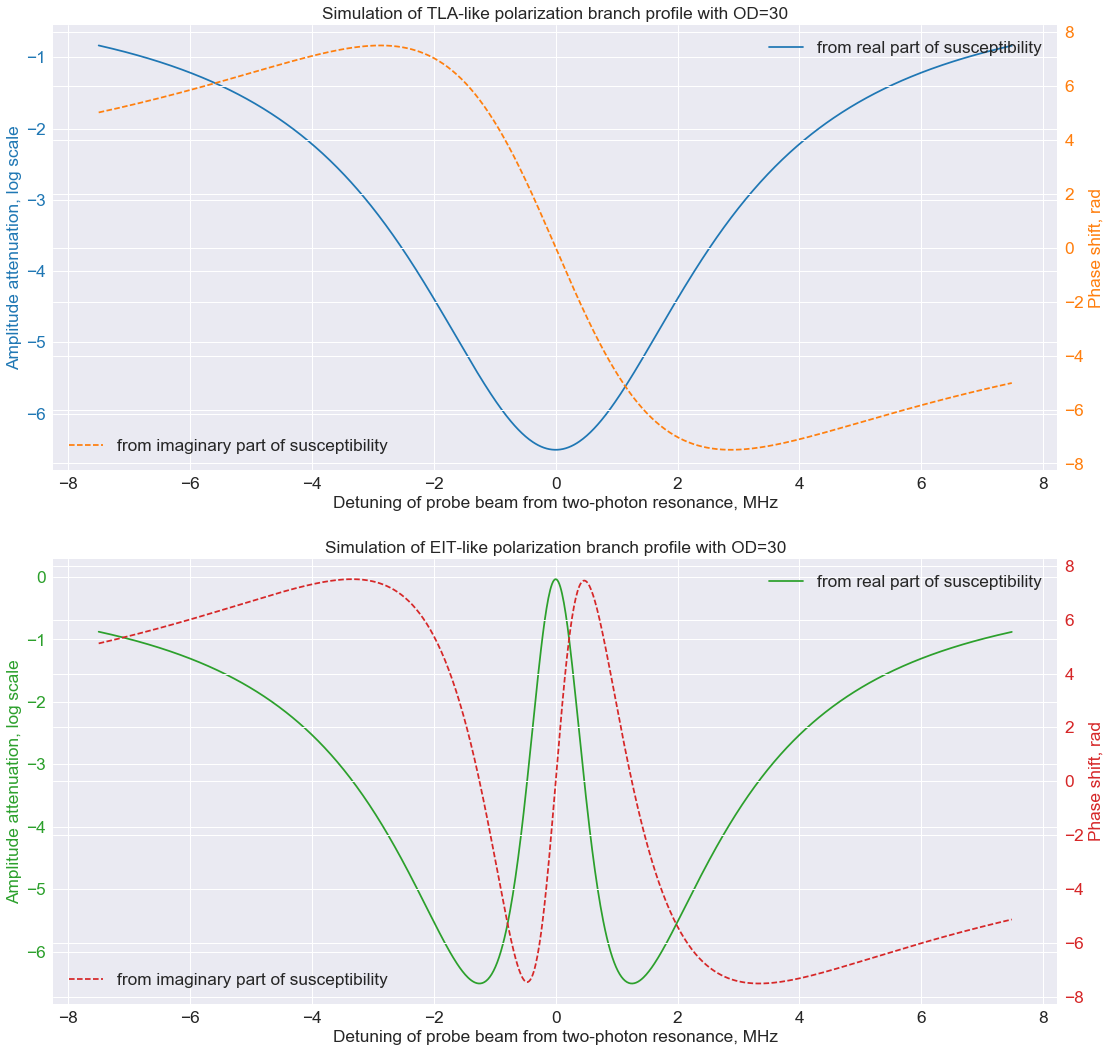}
%left down right up
\end{tabular}
\linespread{1} 
\caption{Numerical simulation for the dispersion of the probe beam after propagation in the optically dense cold atom ensemble (OD = 30), where the atomic structure and coupling laser linkage pattern are according to Fig. \ref{schematic_1}(a). The upper graph is for the TLA-like branch, and the lower graph is for the EIT-like branch. The solid lines are for absorption while the dashed lines are for phase shift. The atomic parameters are $\Gamma = 2\pi\times5.75$MHz which is from $^{87}$Rb D1 transition, $\gamma = 0.001\times\Gamma$; and the optical parameters are $\Omega_c = 2\pi \times 2.5$MHz, $\Delta = 0$.
\label{num_siml_doubleV_1st}}
\end{figure}

From Eq.\eqref{1a_eigenvalues}, an observation can be made that $\lambda_1$ is tied to dynamics similar to a two-level atom (TLA), while $\lambda_2$ is tied to dynamics similar to typical EIT. Upon incidence, the probe light decomposes into two components with opposite polarizations:  $\frac{1}{\sqrt{2}}(\Omega_{p+} + \beta^*_c\Omega_{p-})$ corresponding to the TLA branch, and $\frac{1}{\sqrt{2}}(-\Omega_{p+} + \beta^*_c\Omega_{p-})$ corresponding to the EIT branch. In general, the polarization ingredients for probe light of the two branches are solely determined by the coupling light's polarization. Loosely speaking, the coupling light induces birefringence in the medium such that for the probe laser, the medium is transparent to one  polarization component but opaque to the other. This process can also be equivalently interpreted from the FWM viewpoint. An example of numerical simulation is shown in Fig. \ref{num_siml_doubleV_1st}.
\par
Due to the energy degeneracy of the states $|2\rangle, |3\rangle$, the linkage structure of Fig. \ref{schematic_1}(a) posses a special symmetry from the viewpoint of Morris-Shore transform \cite{PhysRevA.27.906, PhysRevA.74.053402}. Therefore it is necessary to examine the case where this symmetry is broken, namely with the energy degeneracy lifted. This triggers the study of a linkage structure shown in Fig. \ref{schematic_1}(c), where the energies of $|2\rangle$ and $|3\rangle$ differ by $2\hbar\Delta$ and the frequency of the coupling light is naturally chosen to correspond to the energy difference from $|4\rangle$ to the middle of $|2\rangle$ and $|3\rangle$.
\par
Formally, assuming that the probe pulse length $\tau$ is long enough such that $| \tau \cdot \Delta|\gg 1, | \tau \cdot \Omega_c|\gg 1$; the EOM for the system is given by the Maxwell-Bloch equations almost identical to Eq.\eqref{1a_EOM}, with an essential difference of $\Delta \to -\Delta$ in Eq.\eqref{1a_EOM_c}. After applying the steady state condition, the equation governing the dynamics of the probe light's propagation through the medium is: 
\begin{equation}
\label{1c_dynamics_1}
\frac{d}{dz}
\begin{bmatrix}
\Omega_{p+} \\
\Omega_{p-}
\end{bmatrix}
= \frac{i}{2} n \sigma \Gamma B^{-1}
\begin{bmatrix}
\Omega_{p+} \\
\Omega_{p-}
\end{bmatrix}, 
\end{equation}
where $B$ is a 2 by 2 matrix independent of $z$, with $B_{11}=\frac{\Omega^2_c}{2(2\delta+i\gamma)}+2\Delta-2\delta-i\Gamma$, $B_{12} = -\beta^*_c \frac{\Omega^2_c}{2(2\delta+i\gamma)}$, $B_{21} = -\beta_c \frac{\Omega^2_c}{2(2\delta+i\gamma)}$ and $B_{22}=\frac{\Omega^2_c}{2(2\delta+i\gamma)}-2\Delta-2\delta-i\Gamma$.
\par
Eventually the characteristics of Eq.\eqref{1c_dynamics_1} can be studied via analyzing $B$, whose eigenvalues are the following: 
\begin{equation}
\label{1c_dynamics_c}
\lambda_{\pm} = 
\frac{|\Omega_c|^2}{2(2\delta+i\gamma)}-2\delta-i\Gamma
\pm \sqrt{4\Delta^2 + (\frac{|\Omega_c|^2}{2(2\delta+i\gamma)})^2}.
\end{equation}
\par
Analogously, Eq.\eqref{1c_dynamics_1} can be decoupled into two independent branches with different dispersion relations for the incident probe light. Although $\lambda_{\pm}$ are of quite different appearances compared with Eq.\eqref{1a_eigenvalues}, we can still identify that one is associated with a TLA scattering behavior branch while the other one is associated with an EIT scattering behavior branch. To explain this classification, we first examine the situation with $\delta = 0$, where Eq.\eqref{1c_dynamics_c} is reduced to: 
\begin{subequations}
\label{1c_EOM_delta0}
\begin{eqnarray}
\label{1c_EOM_delta0_a}
\frac{d}{dz}(-\beta_c\Omega_{p+} + \Omega_{p-}) = 0, \\
\label{1c_EOM_delta0_b}
\frac{d}{dz}(\beta_c\Omega_{p+} + \Omega_{p-}) = 
-\frac{n\sigma}{2}(\beta_c\Omega_{p+} + \Omega_{p-}).
\end{eqnarray}
\end{subequations}
\begin{figure}[b]
\centering
\begin{tabular}{l}
\includegraphics[width=7.5cm]{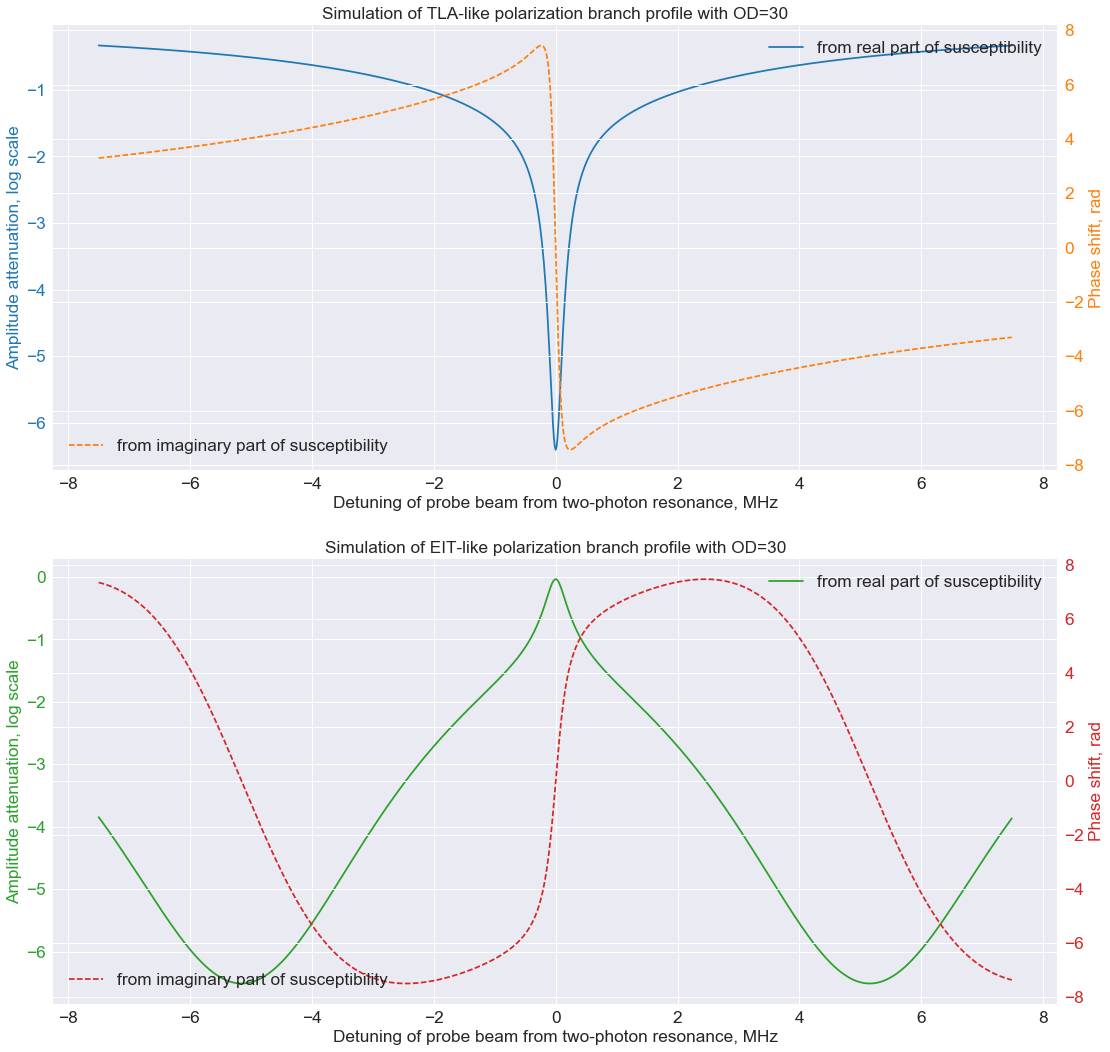}
\end{tabular}
\linespread{1} 
\caption{Numerical simulation for the dispersion of the probe beam after propagation in the optically dense cold atom ensemble (OD = 30), where the atomic structure and coupling laser linkage pattern are according to Fig. \ref{schematic_1}(c). The upper graph is for the TLA-like branch, and the lower graph is for the EIT-like branch. The solid lines are for absorption while the dashed lines are for phase shift. The atomic parameters are $\Gamma = 2\pi\times5.75$MHz which is from $^{87}$Rb D1 transition, $\gamma = 0.001\times\Gamma$, $\Delta = 2\pi\times5$MHz; and the optical parameters are $\Omega_c = 2\pi \times 2.5$MHz. Note that in the neighborhood of $\delta=0$ the behavior is similar to that of Fig. \ref{num_siml_doubleV_1st}.\label{num_siml_doubleV_2nd}}
\end{figure}
For $\delta \neq 0$, we invoke the approximation that $|\delta|$ is small such that $|\Omega_c|^2\gg|2\Delta\cdot4\delta|$ and $\gamma=0$, which leads to $\sqrt{4\Delta^2 + (\frac{|\Omega_c|^2}{4\delta})^2} \approx \frac{|\Omega_c|^2}{4\delta} + \frac{8\delta \Delta^2}{|\Omega_c|^2}$. Then to the lowest order, $\lambda_+ \approx \frac{|\Omega_c|^2}{2\delta} - 2\delta - i\Gamma$ while $\lambda_- \approx -2\delta - i\Gamma$. The two branches' behaviors are EIT-like and TLA-like respectively, and henceforth the previous observation is justified. The numerical simulation is presented in Fig. \ref{num_siml_doubleV_2nd}. In general, if the energy degeneracy of two excited states with different angular momentum is lifted, the polarization decomposition of the incident probe beam is subject to not only the polarization of the coupling light but also its detuning \cite{SuppInfo}, although the signature of polarization induced interference persists. One possible application of an experimental realization of Fig. \ref{schematic_1}(c) is the precision measurement of the relative Zeeman shifts experienced by the excited states.
\par
Ideally, all-optical control and switching in the polarization degree of freedom and polarization filtering can be implemented by utilizing the double-V linkage structure of Fig. \ref{schematic_1}. Nevertheless, it requires the coupling light intensity to be much higher above the single-photon level. If one insists on coupling light of low intensity, then some form of enhancement such as a high finesse optical cavity has to be employed. Otherwise, modifications to the double-V linkage are in need to pursue polarization induced interference within EIT and FWM processes between two weak optical fields.
\par
\begin{figure}[t]
 \centering
\begin{tabular}{l}
\includegraphics[trim = 0mm 0mm 0mm 0mm, clip, width=8.5cm]{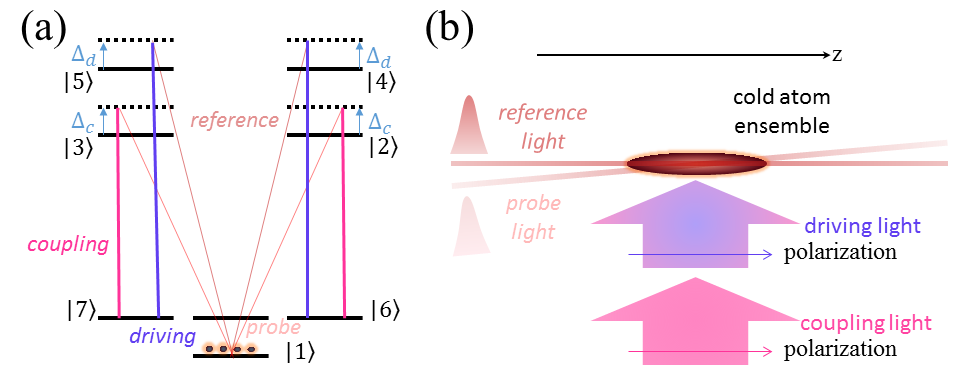}
%left down right up
\end{tabular}
\linespread{1} 
\caption{(a) Schematic of the double-M linkage structure of a four-level seven-state system. It can also be realized in the $^{87}$Rb D1 or D2 transitions \cite{SuppInfo}. (b) Simplified proposal for experimental implementation of the optical configuration. The reference and probe lights are almost parallel with a possible tiny cross angle for the purpose of phase matching. The coupling and driving lasers are assumed to be cw, of the same spatial mode, uniform in intensity across the cold atom ensemble and cross the probe \& reference optical axis at right angle.
\label{schematic_2}}
\end{figure}

In particular, consider the polarization discriminating cross-modulation process of Fig. \ref{schematic_2}, where the linkage structure enables double-EIT for both the $\sigma^+, \sigma^-$ transitions. The coupling and driving lasers are of moderate Rabi frequencies comparable to the linewidth of the transitions, while the incident probe and reference light pulses are of low optical intensities. Then, up to a global phase, the dynamics up to the lowest order can be derived from the Maxwell-Bloch equations \cite{PhysRevLett.115.113005, PhysRevLett.117.203601, PhysRevA.71.011803} in the rotating wave frame, where the polarization is resolved with respect to the quantization axis choice as the z-direction: 
\begin{subequations}
\label{4a_EOM}
\begin{align}
\frac{\partial}{\partial z} \Omega_{p+} + \frac{1}{c} \frac{\partial}{\partial t} \Omega_{p+} = i \frac{n\sigma_p\Gamma_c}{2} \rho_{21};
\\
\frac{\partial}{\partial z} \Omega_{r+} + \frac{1}{c} \frac{\partial}{\partial t} \Omega_{r+} = i \frac{n\sigma_r\Gamma_d}{2} \rho_{41};
\\
\label{1a_EOM_e}
\frac{d}{dt} \rho_{21} = \frac{i}{2}\Omega_{p+} + \frac{i}{2}\Omega_c\rho_{61} + (i\Delta_c - \frac{\Gamma_c}{2})\rho_{21};
\\
\frac{d}{dt} \rho_{41} = \frac{i}{2}\Omega_{r+} + \frac{i}{2}\Omega_d\rho_{61} + (i\Delta_d - \frac{\Gamma_d}{2})\rho_{41};
\\
\frac{d}{dt} \rho_{61} = 
\frac{i}{2}\Omega^*_c\rho_{21}+ \frac{i}{2}\Omega^*_d\rho_{41}
-\frac{\gamma}{2}\rho_{61};
\end{align}
\end{subequations}
where $\Gamma_c$ is the decay rate of states $|2\rangle,|3\rangle$, $\Gamma_d$ is the decay rate of states $|4\rangle, |5\rangle$, $\gamma$ is the decoherence rate of states $|6\rangle, |7\rangle$ which is typically tiny, and $\sigma_{p,r}$ are the atom-light cross sections for the probe and reference lights. The EOMs for $\Omega_{p-}$ and $\Omega_{r-}$ are of the same form with the following replacements: $\rho_{21}\to\rho_{31}, \rho_{41} \to \rho_{51}, \rho_{61} \to \rho_{71}$. 
\par
% talk about symmetry
An inherent symmetry about polarization can be observed in the atom-light interaction with this atomic linkage structure. Eq.\eqref{4a_EOM} is now written with the circular polarization basis, nevertheless its form is invariant under Morris-Shore transform to any orthonormal polarization basis \cite{SuppInfo}. This symmetry guarantees that the polarization induced interference considered here is only up to the relative polarization difference of the probe and reference while no special polarization orientation pre-exists in the system. 
\par
We analyze the dynamics via the steady state solution of Eq.\eqref{4a_EOM} under the assumption of perfect ground level coherence $\gamma=0$ and equivalent detunings $\Delta_c=\Delta_d=\Delta$. Then the dynamics of the probe and reference lights is specified by the following equation:
\begin{equation}
\label{4a_EOM_2}
\frac{d}{dz}
\begin{bmatrix}
\Omega_{p+,-} \\
\Omega_{r+,-} 
\end{bmatrix}
=
\xi \cdot
\underbrace{
\begin{bmatrix}
-\frac{\Omega_d}{\Omega_c}\Gamma_c & \Gamma_c \\
\frac{\Omega_d \Omega^*_c}{\Omega^*_d \Omega_c}\Gamma_d a_r &
-\frac{\Omega^*_c}{\Omega^*_d}\Gamma_d a_r
\end{bmatrix}
}_{M_0}
\begin{bmatrix}
\Omega_{p+,-} \\
\Omega_{r+,-} 
\end{bmatrix},
\end{equation}
with the constants $\xi$ and $a_r$ defined as:
\begin{equation}
\label{4a_prefactor_a}
\xi =
-\frac{n\sigma_p}{4}
\frac{1}{\frac{\Omega_d}{\Omega_c}(i\Delta-\frac{\Gamma_c}{2})
+ \frac{\Omega^*_c}{\Omega^*_d}(i\Delta-\frac{\Gamma_d}{2}) },
\,
a_r = \frac{\sigma_r}{\sigma_p};
\end{equation}
where we can further assume that $\Gamma_c=\Gamma_d=\Gamma$ typically.
\par
For appropriate settings of control parameters, this system can demonstrate cross-modulation capabilities for the probe and reference lights involving the polarization degree of freedoms, provided the optical depth along the propagation axis is adequate. 
\par
In particular, we first analyze the two extreme cases where the polarization states of the probe and reference beams are identical or orthogonal. If their polarization states are identical, the situation reduces to that of a typical FWM, where the cross-modulation is subject to the their initial relative phase difference, just as Refs. \cite{PhysRevLett.115.113005, PhysRevLett.117.203601}. If the polarizations of the incident probe and reference lights are orthogonal $\Omega_{r+}(0)\Omega^*_{p+}(0)+\Omega_{r-}(0)\Omega^*_{p-}(0) = 0$,  then their initial phase difference do not contribute to the their total transmitted power. To observe this, we can simply take a Morris-Shore transform such that in the transformed new polarization basis $\Omega'_{p-}(0) = \Omega'_{r+}(0) = 0$, where $'$ denotes quantities expressed in the new basis.
\par
\begin{figure}[t]
 \centering
\begin{tabular}{l}
\includegraphics[trim = 0mm 0mm 0mm 0mm, clip, width=8.5cm]{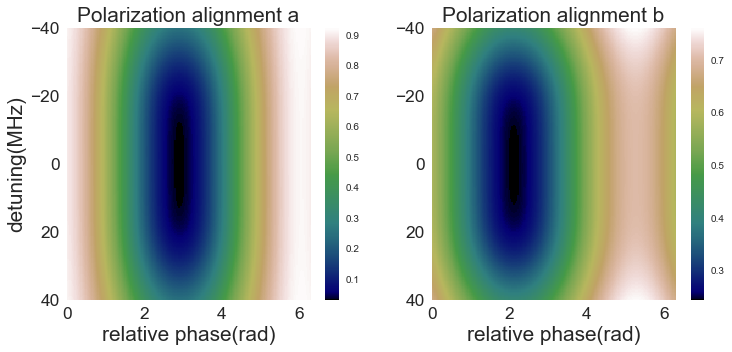}
%left down right up
\end{tabular}
\linespread{1} 
\caption{Numerical simulation for the cross-modulation between probe and reference incidences after propagation in an optically dense cold atom ensemble (OD = 10), where the atomic structure and controlling lasers are set according to Fig. \ref{schematic_2}. The total emergent intensity is plotted with normalization to the total incident intensity, where the incident probe and reference beams are assumed to be linearly polarized and at the same intensity. At alignment a, the angle between two linear polarizations is $\frac{\pi}{6}$, while at alignment b the angle is $\frac{2\pi}{3}$. The relative phase between the two lights is scanned, which is defined with respect to their left circular polarization component. The atomic parameters are $\Gamma_c = \Gamma_d = \Gamma = 2\pi\times6.07$MHz which is from $^{87}$Rb D2 transition, $\gamma = 0.01\times \Gamma$; and the optical parameters are $\Omega_c = \Omega_d = 2\pi \times 5$MHz, where the detuning $\Delta$ is scanned. 
\label{num_siml_doubleM_1st}}
\par
\end{figure}
\begin{figure}[h]
 \centering
\begin{tabular}{l}
\includegraphics[trim = 0mm 0mm 0mm 0mm, clip, width=8.5cm]{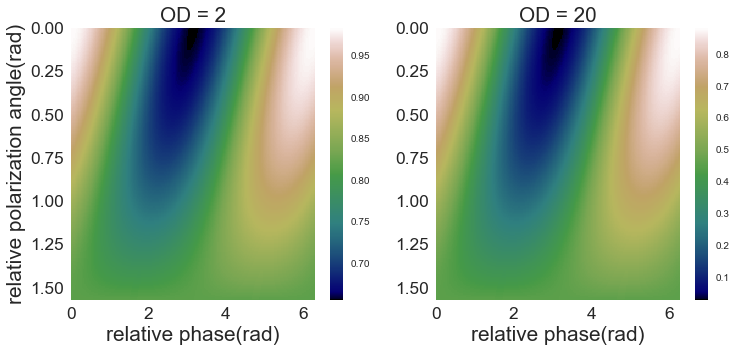}
%left down right up
\end{tabular}
\linespread{1} 
\caption{Numerical simulation for the cross-modulation between probe and reference incidences after propagation in an optically dense cold atom ensemble with OD=2 and 20, where the same atomic parameters and coupling \& driving Rabi frequencies as Fig. \ref{num_siml_doubleM_1st}. The difference is that the detunings are fixed here: $\Delta=5\text{MHz}$ while the relative phase and relative polarization angle is scanned. Note that when the polarizations of the probe and reference lights are orthogonal, the scanning of relative phase does not cause any change in the total emergent intensity. 
\label{num_siml_doubleM_2nd}}
\end{figure}

More specifically, we discuss the cross-modulation of amplitude, and the results of numerical simulations are presented in Figs. \ref{num_siml_doubleM_1st} \& \ref{num_siml_doubleM_2nd}. These numerical simulations are carried out under the assumption that both the incident probe and reference optical fields are linearly polarized. In order to quantify the relative differences in polarization and phase, the incidences with respect to the rotating wave frame are configured as $\Omega_{p+}(0) = \Omega_1, \Omega_{p-}(0) = \Omega_1, \Omega_{r+}(0) = e^{i\theta}\Omega_2, \Omega_{r+}(0) = e^{i\theta}e^{i\varphi}\Omega_2$, where real numbers $\Omega_1, \Omega_2$ are constant amplitude factors, $\theta$ describes the relative phase difference and $\varphi$ describes the relative polarization difference.
\par
To study the dynamics in detail, Eq.\eqref{4a_EOM} can be effectively analyzed by Fourier transform \cite{SuppInfo}. Nevertheless, the steady-state description of Eq.\eqref{4a_EOM_2} suffices to provide insight into the features of the system. In the special on-resonance condition of $\Delta = 0$, the emergent probe light intensity can be computed and it contains a succinct term indicative of interference with clear signature from the polarization:
\begin{equation}
\label{4b_example}
\Re\{
\frac{\Omega_c}{\Omega_d}[\Omega_{r+}(0)\Omega^*_{p+}(0)
+\Omega_{r-}(0)\Omega^*_{p-}(0)]
\},
\end{equation}
where the outcome of interference is up to the relative polarization difference of the probe and reference lights and a phase accumulation involving all four optical fields. More generally, Eq.\eqref{4a_EOM_2} permits two modes of different dynamics with respect to the two eigenvalues of $M_0$,
\begin{equation}
\label{4a_eigenvalues}
\lambda_1 = 0,\,
\lambda_2 = (-\frac{\Omega_d}{\Omega_c} - \frac{\Omega^*_c}{\Omega^*_d})\Gamma;
\end{equation}
where $\lambda_1$ corresponds to an EIT-like transparency window and $\lambda_2$ corresponds to a TLA-like dispersion. 
There exists non-trivial $\mathbf{e} = \begin{bmatrix}e_1\\e_2\end{bmatrix}$ for $\lambda_1$ such that $\frac{d}{dz} (e^*_1 \Omega_{p\pm} + e^*_2 \Omega_{r\pm}) = 0$. In other words, for the right circularly polarized components of the incidences $\mathbf{\Omega}_+(0) = \begin{bmatrix}\Omega_{p+}(0)\\ \Omega_{r+}(0)\end{bmatrix}$, it projection onto $\mathbf{e}$ is subject to the transparency, which is the same case for the left circularly polarized components of the incidences $\mathbf{\Omega}_-(0) = \begin{bmatrix}\Omega_{p-}(0)\\ \Omega_{r-}(0)\end{bmatrix}$. In particular, we find that both the probe and reference frequency components for the transparency window share the same polarization. 
\par
\begin{figure}[t]
 \centering
\begin{tabular}{l}
\includegraphics[trim = 0mm 0mm 0mm 0mm, clip, width=7.5cm]{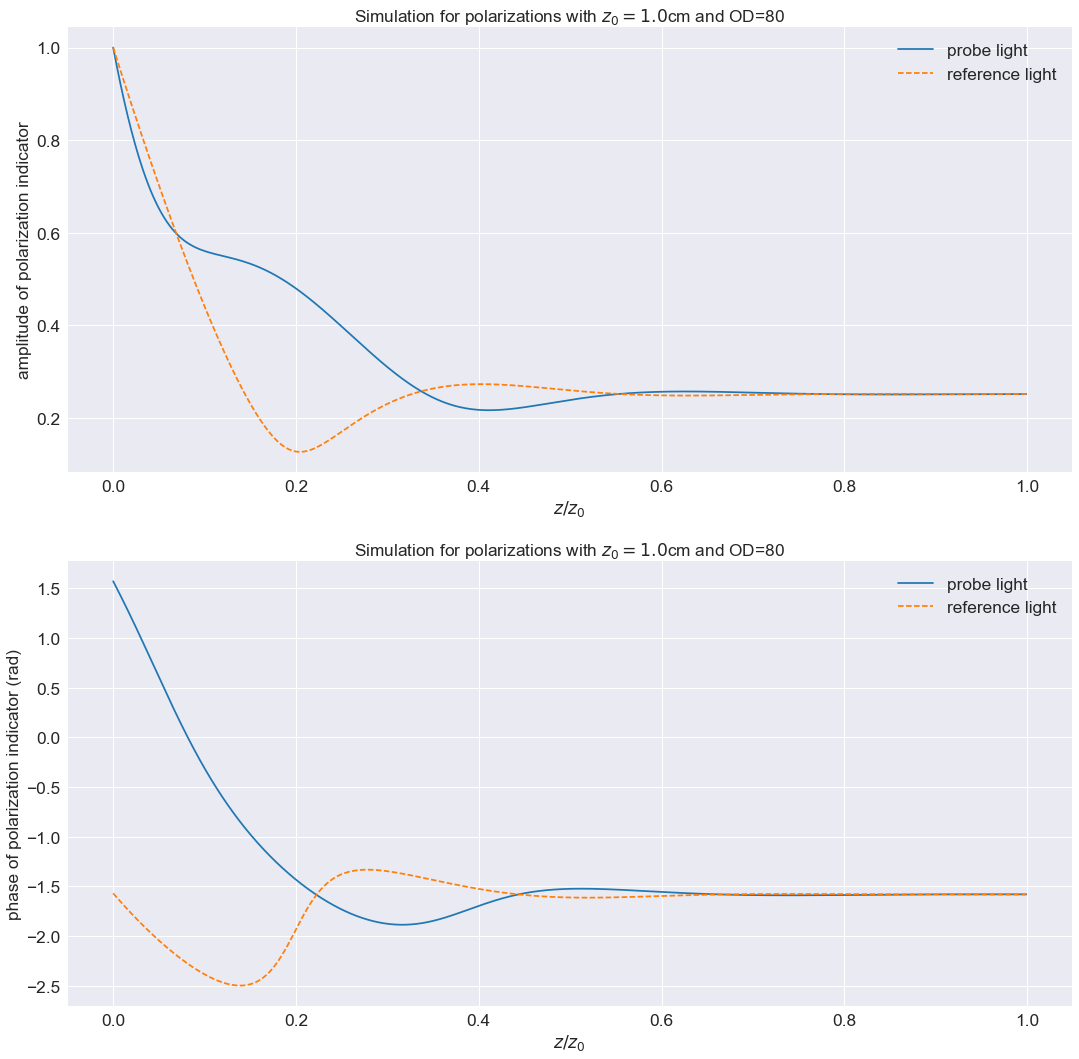}
%left down right up
\end{tabular}
\linespread{1} 
\caption{Numerical simulation for the polarization dynamics during propagation, where the cold atom ensemble is assumed to be uniformly distributed, of length $z_0 = 1$cm and OD=80. The polarization indicator $\zeta$ for a general polarization $P$ is defined as $P = \kappa( P_+ + \zeta P_-), \text{ with } \kappa \equiv (1 + |\zeta|^2)^{-\frac{1}{2}}$ where $P_{+,-}$ are the right and left circular polarization basis.
The atomic parameters are $\Gamma = 2\pi\times6.07$MHz associated with $^{87}$Rb D2 transition, $\gamma = 0.01\times \Gamma$; and the optical parameters are $\Omega_c = \Omega_d = 2\pi \times 5$MHz, $\Delta=5\text{MHz}$, together with the incidences $\Omega_{p+}(0) = \Omega_1, \Omega_{p-}(0) = \Omega_1, \Omega_{r+}(0) = i\Omega_1, \Omega_{r+}(0) = -i\Omega_1$ where $\Omega_1$ is a common constant amplitude factor which eventually cancels out.
\label{num_siml_doubleM_3rd}}
\end{figure}

The above observation sketches the inherent characteristics of such a system. Upon the probe and reference incidences of some prescribed initial polarizations and phases,  they immediately recombine into two composite pulses corresponding to two modes of propagation dynamics respectively. Each mode contains both the probe and reference frequency components of identical polarizations up to a phase difference. One mode experiences EIT-like transparency while the other mode experience TLA-like dispersion, with different group velocities. Usually, the TLA-like mode decays away rather rapidly, whose energy is dissipated via spontaneous emission, and therefore only one mode of composite pulses is emergent from the optically thick cold atom medium. A numerical simulation of the polarization change with the propagation along the z-direction is given in Fig. \ref{num_siml_doubleM_3rd}.
\par
\par
The derivations so far are essentially within the semi-classical framework, where the probe and reference incidences are assumed to be weak coherent classical optical pulses. Even though the analysis holds for very low incident power, the natural question to ask is whether it makes sense for genuine single-photon incidences. In principle, due to the fact that the EOM Eq.\eqref{4a_EOM} is linear in $\Omega_{p\pm}, \Omega_{r\pm}$ \cite{RevModPhys.82.1041}, it is anticipated that the polarization induced interference still exists if both the probe and reference optical fields are quantized. The detailed analysis of this issue is an interesting subject for future work.
\par
%Conclusion
In conclusion, we have proposed particular forms of atomic linkage structures where the polarization of the optical fields is of essential role in the atom-light interaction. We have shown that the polarization induced quantum interference comes naturally within the EIT and FWM processes from the double-V and double-M atomic linkage structures. We have also studied the fundamental properties of the polarization-dependent cross-modulation in the double-M structure. We hope that our work helps the effort of realizing strong polarization-dependent non-linear interactions between weak optical pulses down to the single-photon level in cold atom medium. We also hope that it helps the research into the polarization degree of freedom from quantum optics perspective on the topic of stimulated Raman adiabatic passage \cite{PhysRevA.90.033408}.

\begin{acknowledgments}
The authors acknowledge support by the National Key R\&D Program of China (Grant No. 2016YFA0301504). 
The authors gratefully acknowledge Professor Harold Metcalf who essentially initiated and prompted this work. The authors thank Professor Mark Saffman for carefully reviewing the manuscript and enlightening discussions.
\end{acknowledgments}

\bibliographystyle{apsrev4-1}

\renewcommand{\baselinestretch}{1}
\normalsize

%\clearpage%
%\phantomsection%
%\addcontentsline{toc}{chapter}{\numberline{}{Bibliography}}%
\bibliography{ysref}

\begin{thebibliography}{35}%
\makeatletter
\providecommand \@ifxundefined [1]{%
 \@ifx{#1\undefined}
}%
\providecommand \@ifnum [1]{%
 \ifnum #1\expandafter \@firstoftwo
 \else \expandafter \@secondoftwo
 \fi
}%
\providecommand \@ifx [1]{%
 \ifx #1\expandafter \@firstoftwo
 \else \expandafter \@secondoftwo
 \fi
}%
\providecommand \natexlab [1]{#1}%
\providecommand \enquote  [1]{``#1''}%
\providecommand \bibnamefont  [1]{#1}%
\providecommand \bibfnamefont [1]{#1}%
\providecommand \citenamefont [1]{#1}%
\providecommand \href@noop [0]{\@secondoftwo}%
\providecommand \href [0]{\begingroup \@sanitize@url \@href}%
\providecommand \@href[1]{\@@startlink{#1}\@@href}%
\providecommand \@@href[1]{\endgroup#1\@@endlink}%
\providecommand \@sanitize@url [0]{\catcode `\\12\catcode `\$12\catcode
  `\&12\catcode `\#12\catcode `\^12\catcode `\_12\catcode `\%12\relax}%
\providecommand \@@startlink[1]{}%
\providecommand \@@endlink[0]{}%
\providecommand \url  [0]{\begingroup\@sanitize@url \@url }%
\providecommand \@url [1]{\endgroup\@href {#1}{\urlprefix }}%
\providecommand \urlprefix  [0]{URL }%
\providecommand \Eprint [0]{\href }%
\providecommand \doibase [0]{http://dx.doi.org/}%
\providecommand \selectlanguage [0]{\@gobble}%
\providecommand \bibinfo  [0]{\@secondoftwo}%
\providecommand \bibfield  [0]{\@secondoftwo}%
\providecommand \translation [1]{[#1]}%
\providecommand \BibitemOpen [0]{}%
\providecommand \bibitemStop [0]{}%
\providecommand \bibitemNoStop [0]{.\EOS\space}%
\providecommand \EOS [0]{\spacefactor3000\relax}%
\providecommand \BibitemShut  [1]{\csname bibitem#1\endcsname}%
\let\auto@bib@innerbib\@empty
%</preamble>
\bibitem [{\citenamefont {Ling}\ \emph {et~al.}(1996)\citenamefont {Ling},
  \citenamefont {Li},\ and\ \citenamefont {Xiao}}]{PhysRevA.53.1014}%
  \BibitemOpen
  \bibfield  {author} {\bibinfo {author} {\bibfnamefont {H.~Y.}\ \bibnamefont
  {Ling}}, \bibinfo {author} {\bibfnamefont {Y.-Q.}\ \bibnamefont {Li}}, \ and\
  \bibinfo {author} {\bibfnamefont {M.}~\bibnamefont {Xiao}},\ }\href {\doibase
  10.1103/PhysRevA.53.1014} {\bibfield  {journal} {\bibinfo  {journal} {Phys.
  Rev. A}\ }\textbf {\bibinfo {volume} {53}},\ \bibinfo {pages} {1014}
  (\bibinfo {year} {1996})}\BibitemShut {NoStop}%
\bibitem [{\citenamefont {Chen}\ \emph {et~al.}(2000)\citenamefont {Chen},
  \citenamefont {Lin},\ and\ \citenamefont {Yu}}]{PhysRevA.61.053805}%
  \BibitemOpen
  \bibfield  {author} {\bibinfo {author} {\bibfnamefont {Y.-C.}\ \bibnamefont
  {Chen}}, \bibinfo {author} {\bibfnamefont {C.-W.}\ \bibnamefont {Lin}}, \
  and\ \bibinfo {author} {\bibfnamefont {I.~A.}\ \bibnamefont {Yu}},\ }\href
  {\doibase 10.1103/PhysRevA.61.053805} {\bibfield  {journal} {\bibinfo
  {journal} {Phys. Rev. A}\ }\textbf {\bibinfo {volume} {61}},\ \bibinfo
  {pages} {053805} (\bibinfo {year} {2000})}\BibitemShut {NoStop}%
\bibitem [{\citenamefont {Magno}\ \emph {et~al.}(2001)\citenamefont {Magno},
  \citenamefont {Prandini}, \citenamefont {Nussenzveig},\ and\ \citenamefont
  {Vianna}}]{PhysRevA.63.063406}%
  \BibitemOpen
  \bibfield  {author} {\bibinfo {author} {\bibfnamefont {W.~C.}\ \bibnamefont
  {Magno}}, \bibinfo {author} {\bibfnamefont {R.~B.}\ \bibnamefont {Prandini}},
  \bibinfo {author} {\bibfnamefont {P.}~\bibnamefont {Nussenzveig}}, \ and\
  \bibinfo {author} {\bibfnamefont {S.~S.}\ \bibnamefont {Vianna}},\ }\href
  {\doibase 10.1103/PhysRevA.63.063406} {\bibfield  {journal} {\bibinfo
  {journal} {Phys. Rev. A}\ }\textbf {\bibinfo {volume} {63}},\ \bibinfo
  {pages} {063406} (\bibinfo {year} {2001})}\BibitemShut {NoStop}%
\bibitem [{\citenamefont {Li}\ \emph {et~al.}(2010)\citenamefont {Li},
  \citenamefont {Zhang}, \citenamefont {Nie}, \citenamefont {Du}, \citenamefont
  {Wang}, \citenamefont {Song},\ and\ \citenamefont
  {Xiao}}]{PhysRevA.81.033801}%
  \BibitemOpen
  \bibfield  {author} {\bibinfo {author} {\bibfnamefont {C.}~\bibnamefont
  {Li}}, \bibinfo {author} {\bibfnamefont {Y.}~\bibnamefont {Zhang}}, \bibinfo
  {author} {\bibfnamefont {Z.}~\bibnamefont {Nie}}, \bibinfo {author}
  {\bibfnamefont {Y.}~\bibnamefont {Du}}, \bibinfo {author} {\bibfnamefont
  {R.}~\bibnamefont {Wang}}, \bibinfo {author} {\bibfnamefont {J.}~\bibnamefont
  {Song}}, \ and\ \bibinfo {author} {\bibfnamefont {M.}~\bibnamefont {Xiao}},\
  }\href {\doibase 10.1103/PhysRevA.81.033801} {\bibfield  {journal} {\bibinfo
  {journal} {Phys. Rev. A}\ }\textbf {\bibinfo {volume} {81}},\ \bibinfo
  {pages} {033801} (\bibinfo {year} {2010})}\BibitemShut {NoStop}%
\bibitem [{\citenamefont {Reshetov}\ and\ \citenamefont
  {Meleshko}(2014)}]{LaserPhys.24.094011}%
  \BibitemOpen
  \bibfield  {author} {\bibinfo {author} {\bibfnamefont {V.~A.}\ \bibnamefont
  {Reshetov}}\ and\ \bibinfo {author} {\bibfnamefont {I.~V.}\ \bibnamefont
  {Meleshko}},\ }\href {\doibase 10.1088/1054-660X/24/9/094011} {\bibfield
  {journal} {\bibinfo  {journal} {Laser Phys.}\ }\textbf {\bibinfo {volume}
  {24}},\ \bibinfo {pages} {094011} (\bibinfo {year} {2014})}\BibitemShut
  {NoStop}%
\bibitem [{\citenamefont {Bao}\ \emph {et~al.}(2016)\citenamefont {Bao},
  \citenamefont {Zhang}, \citenamefont {Zhou}, \citenamefont {Zhang},
  \citenamefont {Zhao}, \citenamefont {Xiao},\ and\ \citenamefont
  {Jia}}]{PhysRevA.94.043822}%
  \BibitemOpen
  \bibfield  {author} {\bibinfo {author} {\bibfnamefont {S.}~\bibnamefont
  {Bao}}, \bibinfo {author} {\bibfnamefont {H.}~\bibnamefont {Zhang}}, \bibinfo
  {author} {\bibfnamefont {J.}~\bibnamefont {Zhou}}, \bibinfo {author}
  {\bibfnamefont {L.}~\bibnamefont {Zhang}}, \bibinfo {author} {\bibfnamefont
  {J.}~\bibnamefont {Zhao}}, \bibinfo {author} {\bibfnamefont {L.}~\bibnamefont
  {Xiao}}, \ and\ \bibinfo {author} {\bibfnamefont {S.}~\bibnamefont {Jia}},\
  }\href {\doibase 10.1103/PhysRevA.94.043822} {\bibfield  {journal} {\bibinfo
  {journal} {Phys. Rev. A}\ }\textbf {\bibinfo {volume} {94}},\ \bibinfo
  {pages} {043822} (\bibinfo {year} {2016})}\BibitemShut {NoStop}%
\bibitem [{\citenamefont {Morris}\ and\ \citenamefont
  {Shore}(1983)}]{PhysRevA.27.906}%
  \BibitemOpen
  \bibfield  {author} {\bibinfo {author} {\bibfnamefont {J.~R.}\ \bibnamefont
  {Morris}}\ and\ \bibinfo {author} {\bibfnamefont {B.~W.}\ \bibnamefont
  {Shore}},\ }\href {\doibase 10.1103/PhysRevA.27.906} {\bibfield  {journal}
  {\bibinfo  {journal} {Phys. Rev. A}\ }\textbf {\bibinfo {volume} {27}},\
  \bibinfo {pages} {906} (\bibinfo {year} {1983})}\BibitemShut {NoStop}%
\bibitem [{\citenamefont {Rangelov}\ \emph {et~al.}(2006)\citenamefont
  {Rangelov}, \citenamefont {Vitanov},\ and\ \citenamefont
  {Shore}}]{PhysRevA.74.053402}%
  \BibitemOpen
  \bibfield  {author} {\bibinfo {author} {\bibfnamefont {A.~A.}\ \bibnamefont
  {Rangelov}}, \bibinfo {author} {\bibfnamefont {N.~V.}\ \bibnamefont
  {Vitanov}}, \ and\ \bibinfo {author} {\bibfnamefont {B.~W.}\ \bibnamefont
  {Shore}},\ }\href {\doibase 10.1103/PhysRevA.74.053402} {\bibfield  {journal}
  {\bibinfo  {journal} {Phys. Rev. A}\ }\textbf {\bibinfo {volume} {74}},\
  \bibinfo {pages} {053402} (\bibinfo {year} {2006})}\BibitemShut {NoStop}%
\bibitem [{\citenamefont {Guan}\ and\ \citenamefont
  {Yu}(2007)}]{PhysRevA.76.033817}%
  \BibitemOpen
  \bibfield  {author} {\bibinfo {author} {\bibfnamefont {P.-C.}\ \bibnamefont
  {Guan}}\ and\ \bibinfo {author} {\bibfnamefont {I.~A.}\ \bibnamefont {Yu}},\
  }\href {\doibase 10.1103/PhysRevA.76.033817} {\bibfield  {journal} {\bibinfo
  {journal} {Phys. Rev. A}\ }\textbf {\bibinfo {volume} {76}},\ \bibinfo
  {pages} {033817} (\bibinfo {year} {2007})}\BibitemShut {NoStop}%
\bibitem [{\citenamefont {Li}\ \emph {et~al.}(2006)\citenamefont {Li},
  \citenamefont {Wang}, \citenamefont {Yang}, \citenamefont {Han},
  \citenamefont {Wang}, \citenamefont {Xiao},\ and\ \citenamefont
  {Peng}}]{PhysRevA.74.033821}%
  \BibitemOpen
  \bibfield  {author} {\bibinfo {author} {\bibfnamefont {S.}~\bibnamefont
  {Li}}, \bibinfo {author} {\bibfnamefont {B.}~\bibnamefont {Wang}}, \bibinfo
  {author} {\bibfnamefont {X.}~\bibnamefont {Yang}}, \bibinfo {author}
  {\bibfnamefont {Y.}~\bibnamefont {Han}}, \bibinfo {author} {\bibfnamefont
  {H.}~\bibnamefont {Wang}}, \bibinfo {author} {\bibfnamefont {M.}~\bibnamefont
  {Xiao}}, \ and\ \bibinfo {author} {\bibfnamefont {K.~C.}\ \bibnamefont
  {Peng}},\ }\href {\doibase 10.1103/PhysRevA.74.033821} {\bibfield  {journal}
  {\bibinfo  {journal} {Phys. Rev. A}\ }\textbf {\bibinfo {volume} {74}},\
  \bibinfo {pages} {033821} (\bibinfo {year} {2006})}\BibitemShut {NoStop}%
\bibitem [{\citenamefont {Hashmi}\ and\ \citenamefont
  {Bouchene}(2008)}]{PhysRevA.77.051803}%
  \BibitemOpen
  \bibfield  {author} {\bibinfo {author} {\bibfnamefont {F.~A.}\ \bibnamefont
  {Hashmi}}\ and\ \bibinfo {author} {\bibfnamefont {M.~A.}\ \bibnamefont
  {Bouchene}},\ }\href {\doibase 10.1103/PhysRevA.77.051803} {\bibfield
  {journal} {\bibinfo  {journal} {Phys. Rev. A}\ }\textbf {\bibinfo {volume}
  {77}},\ \bibinfo {pages} {051803} (\bibinfo {year} {2008})}\BibitemShut
  {NoStop}%
\bibitem [{\citenamefont {Laupr\^etre}\ \emph {et~al.}(2012)\citenamefont
  {Laupr\^etre}, \citenamefont {Kumar}, \citenamefont {Berger}, \citenamefont
  {Faoro}, \citenamefont {Ghosh}, \citenamefont {Bretenaker},\ and\
  \citenamefont {Goldfarb}}]{PhysRevA.85.051805}%
  \BibitemOpen
  \bibfield  {author} {\bibinfo {author} {\bibfnamefont {T.}~\bibnamefont
  {Laupr\^etre}}, \bibinfo {author} {\bibfnamefont {S.}~\bibnamefont {Kumar}},
  \bibinfo {author} {\bibfnamefont {P.}~\bibnamefont {Berger}}, \bibinfo
  {author} {\bibfnamefont {R.}~\bibnamefont {Faoro}}, \bibinfo {author}
  {\bibfnamefont {R.}~\bibnamefont {Ghosh}}, \bibinfo {author} {\bibfnamefont
  {F.}~\bibnamefont {Bretenaker}}, \ and\ \bibinfo {author} {\bibfnamefont
  {F.}~\bibnamefont {Goldfarb}},\ }\href {\doibase 10.1103/PhysRevA.85.051805}
  {\bibfield  {journal} {\bibinfo  {journal} {Phys. Rev. A}\ }\textbf {\bibinfo
  {volume} {85}},\ \bibinfo {pages} {051805} (\bibinfo {year}
  {2012})}\BibitemShut {NoStop}%
\bibitem [{\citenamefont {Zheng}\ \emph {et~al.}(2009)\citenamefont {Zheng},
  \citenamefont {Zhang}, \citenamefont {Khadka}, \citenamefont {Wang},
  \citenamefont {Li}, \citenamefont {Nie},\ and\ \citenamefont
  {Xiao}}]{Zheng:09_MinXiao_OE}%
  \BibitemOpen
  \bibfield  {author} {\bibinfo {author} {\bibfnamefont {H.}~\bibnamefont
  {Zheng}}, \bibinfo {author} {\bibfnamefont {Y.}~\bibnamefont {Zhang}},
  \bibinfo {author} {\bibfnamefont {U.}~\bibnamefont {Khadka}}, \bibinfo
  {author} {\bibfnamefont {R.}~\bibnamefont {Wang}}, \bibinfo {author}
  {\bibfnamefont {C.}~\bibnamefont {Li}}, \bibinfo {author} {\bibfnamefont
  {Z.}~\bibnamefont {Nie}}, \ and\ \bibinfo {author} {\bibfnamefont
  {M.}~\bibnamefont {Xiao}},\ }\href {\doibase 10.1364/OE.17.015468} {\bibfield
   {journal} {\bibinfo  {journal} {Opt. Express}\ }\textbf {\bibinfo {volume}
  {17}},\ \bibinfo {pages} {15468} (\bibinfo {year} {2009})}\BibitemShut
  {NoStop}%
\bibitem [{\citenamefont {Khadka}\ \emph {et~al.}(2010)\citenamefont {Khadka},
  \citenamefont {Zhang},\ and\ \citenamefont {Xiao}}]{PhysRevA.81.023830}%
  \BibitemOpen
  \bibfield  {author} {\bibinfo {author} {\bibfnamefont {U.}~\bibnamefont
  {Khadka}}, \bibinfo {author} {\bibfnamefont {Y.}~\bibnamefont {Zhang}}, \
  and\ \bibinfo {author} {\bibfnamefont {M.}~\bibnamefont {Xiao}},\ }\href
  {\doibase 10.1103/PhysRevA.81.023830} {\bibfield  {journal} {\bibinfo
  {journal} {Phys. Rev. A}\ }\textbf {\bibinfo {volume} {81}},\ \bibinfo
  {pages} {023830} (\bibinfo {year} {2010})}\BibitemShut {NoStop}%
\bibitem [{\citenamefont {Yudin}\ \emph {et~al.}(2010)\citenamefont {Yudin},
  \citenamefont {Taichenachev}, \citenamefont {Dudin}, \citenamefont
  {Velichansky}, \citenamefont {Zibrov},\ and\ \citenamefont
  {Zibrov}}]{PhysRevA.82.033807}%
  \BibitemOpen
  \bibfield  {author} {\bibinfo {author} {\bibfnamefont {V.~I.}\ \bibnamefont
  {Yudin}}, \bibinfo {author} {\bibfnamefont {A.~V.}\ \bibnamefont
  {Taichenachev}}, \bibinfo {author} {\bibfnamefont {Y.~O.}\ \bibnamefont
  {Dudin}}, \bibinfo {author} {\bibfnamefont {V.~L.}\ \bibnamefont
  {Velichansky}}, \bibinfo {author} {\bibfnamefont {A.~S.}\ \bibnamefont
  {Zibrov}}, \ and\ \bibinfo {author} {\bibfnamefont {S.~A.}\ \bibnamefont
  {Zibrov}},\ }\href {\doibase 10.1103/PhysRevA.82.033807} {\bibfield
  {journal} {\bibinfo  {journal} {Phys. Rev. A}\ }\textbf {\bibinfo {volume}
  {82}},\ \bibinfo {pages} {033807} (\bibinfo {year} {2010})}\BibitemShut
  {NoStop}%
\bibitem [{\citenamefont {Tanji}\ \emph {et~al.}(2009)\citenamefont {Tanji},
  \citenamefont {Ghosh}, \citenamefont {Simon}, \citenamefont {Bloom},\ and\
  \citenamefont {Vuleti\ifmmode~\acute{c}\else
  \'{c}\fi{}}}]{PhysRevLett.103.043601}%
  \BibitemOpen
  \bibfield  {author} {\bibinfo {author} {\bibfnamefont {H.}~\bibnamefont
  {Tanji}}, \bibinfo {author} {\bibfnamefont {S.}~\bibnamefont {Ghosh}},
  \bibinfo {author} {\bibfnamefont {J.}~\bibnamefont {Simon}}, \bibinfo
  {author} {\bibfnamefont {B.}~\bibnamefont {Bloom}}, \ and\ \bibinfo {author}
  {\bibfnamefont {V.}~\bibnamefont {Vuleti\ifmmode~\acute{c}\else
  \'{c}\fi{}}},\ }\href {\doibase 10.1103/PhysRevLett.103.043601} {\bibfield
  {journal} {\bibinfo  {journal} {Phys. Rev. Lett.}\ }\textbf {\bibinfo
  {volume} {103}},\ \bibinfo {pages} {043601} (\bibinfo {year}
  {2009})}\BibitemShut {NoStop}%
\bibitem [{\citenamefont {Xu}\ \emph {et~al.}(2013)\citenamefont {Xu},
  \citenamefont {Wu}, \citenamefont {Tian}, \citenamefont {Chen}, \citenamefont
  {Zhang}, \citenamefont {Yan}, \citenamefont {Li}, \citenamefont {Wang},
  \citenamefont {Xie},\ and\ \citenamefont {Peng}}]{PhysRevLett.111.240503}%
  \BibitemOpen
  \bibfield  {author} {\bibinfo {author} {\bibfnamefont {Z.}~\bibnamefont
  {Xu}}, \bibinfo {author} {\bibfnamefont {Y.}~\bibnamefont {Wu}}, \bibinfo
  {author} {\bibfnamefont {L.}~\bibnamefont {Tian}}, \bibinfo {author}
  {\bibfnamefont {L.}~\bibnamefont {Chen}}, \bibinfo {author} {\bibfnamefont
  {Z.}~\bibnamefont {Zhang}}, \bibinfo {author} {\bibfnamefont
  {Z.}~\bibnamefont {Yan}}, \bibinfo {author} {\bibfnamefont {S.}~\bibnamefont
  {Li}}, \bibinfo {author} {\bibfnamefont {H.}~\bibnamefont {Wang}}, \bibinfo
  {author} {\bibfnamefont {C.}~\bibnamefont {Xie}}, \ and\ \bibinfo {author}
  {\bibfnamefont {K.}~\bibnamefont {Peng}},\ }\href {\doibase
  10.1103/PhysRevLett.111.240503} {\bibfield  {journal} {\bibinfo  {journal}
  {Phys. Rev. Lett.}\ }\textbf {\bibinfo {volume} {111}},\ \bibinfo {pages}
  {240503} (\bibinfo {year} {2013})}\BibitemShut {NoStop}%
\bibitem [{\citenamefont {Lukin}\ and\ \citenamefont
  {Imamo\ifmmode~\breve{g}\else \u{g}\fi{}lu}(2000)}]{PhysRevLett.84.1419}%
  \BibitemOpen
  \bibfield  {author} {\bibinfo {author} {\bibfnamefont {M.~D.}\ \bibnamefont
  {Lukin}}\ and\ \bibinfo {author} {\bibfnamefont {A.}~\bibnamefont
  {Imamo\ifmmode~\breve{g}\else \u{g}\fi{}lu}},\ }\href {\doibase
  10.1103/PhysRevLett.84.1419} {\bibfield  {journal} {\bibinfo  {journal}
  {Phys. Rev. Lett.}\ }\textbf {\bibinfo {volume} {84}},\ \bibinfo {pages}
  {1419} (\bibinfo {year} {2000})}\BibitemShut {NoStop}%
\bibitem [{\citenamefont {Harris}\ and\ \citenamefont
  {Hau}(1999)}]{PhysRevLett.82.4611}%
  \BibitemOpen
  \bibfield  {author} {\bibinfo {author} {\bibfnamefont {S.~E.}\ \bibnamefont
  {Harris}}\ and\ \bibinfo {author} {\bibfnamefont {L.~V.}\ \bibnamefont
  {Hau}},\ }\href {\doibase 10.1103/PhysRevLett.82.4611} {\bibfield  {journal}
  {\bibinfo  {journal} {Phys. Rev. Lett.}\ }\textbf {\bibinfo {volume} {82}},\
  \bibinfo {pages} {4611} (\bibinfo {year} {1999})}\BibitemShut {NoStop}%
\bibitem [{\citenamefont {Turchette}\ \emph {et~al.}(1995)\citenamefont
  {Turchette}, \citenamefont {Hood}, \citenamefont {Lange}, \citenamefont
  {Mabuchi},\ and\ \citenamefont {Kimble}}]{PhysRevLett.75.4710}%
  \BibitemOpen
  \bibfield  {author} {\bibinfo {author} {\bibfnamefont {Q.~A.}\ \bibnamefont
  {Turchette}}, \bibinfo {author} {\bibfnamefont {C.~J.}\ \bibnamefont {Hood}},
  \bibinfo {author} {\bibfnamefont {W.}~\bibnamefont {Lange}}, \bibinfo
  {author} {\bibfnamefont {H.}~\bibnamefont {Mabuchi}}, \ and\ \bibinfo
  {author} {\bibfnamefont {H.~J.}\ \bibnamefont {Kimble}},\ }\href {\doibase
  10.1103/PhysRevLett.75.4710} {\bibfield  {journal} {\bibinfo  {journal}
  {Phys. Rev. Lett.}\ }\textbf {\bibinfo {volume} {75}},\ \bibinfo {pages}
  {4710} (\bibinfo {year} {1995})}\BibitemShut {NoStop}%
\bibitem [{\citenamefont {Tanji-Suzuki}\ \emph {et~al.}(2011)\citenamefont
  {Tanji-Suzuki}, \citenamefont {Chen}, \citenamefont {Landig}, \citenamefont
  {Simon},\ and\ \citenamefont {Vuleti{\'c}}}]{VIT_science1266}%
  \BibitemOpen
  \bibfield  {author} {\bibinfo {author} {\bibfnamefont {H.}~\bibnamefont
  {Tanji-Suzuki}}, \bibinfo {author} {\bibfnamefont {W.}~\bibnamefont {Chen}},
  \bibinfo {author} {\bibfnamefont {R.}~\bibnamefont {Landig}}, \bibinfo
  {author} {\bibfnamefont {J.}~\bibnamefont {Simon}}, \ and\ \bibinfo {author}
  {\bibfnamefont {V.}~\bibnamefont {Vuleti{\'c}}},\ }\href {\doibase
  10.1126/science.1208066} {\bibfield  {journal} {\bibinfo  {journal}
  {Science}\ }\textbf {\bibinfo {volume} {333}},\ \bibinfo {pages} {1266}
  (\bibinfo {year} {2011})}\BibitemShut {NoStop}%
\bibitem [{\citenamefont {Reiserer}\ and\ \citenamefont
  {Rempe}(2015)}]{RevModPhys.87.1379}%
  \BibitemOpen
  \bibfield  {author} {\bibinfo {author} {\bibfnamefont {A.}~\bibnamefont
  {Reiserer}}\ and\ \bibinfo {author} {\bibfnamefont {G.}~\bibnamefont
  {Rempe}},\ }\href {\doibase 10.1103/RevModPhys.87.1379} {\bibfield  {journal}
  {\bibinfo  {journal} {Rev. Mod. Phys.}\ }\textbf {\bibinfo {volume} {87}},\
  \bibinfo {pages} {1379} (\bibinfo {year} {2015})}\BibitemShut {NoStop}%
\bibitem [{\citenamefont {Gorshkov}\ \emph {et~al.}(2011)\citenamefont
  {Gorshkov}, \citenamefont {Otterbach}, \citenamefont {Fleischhauer},
  \citenamefont {Pohl},\ and\ \citenamefont {Lukin}}]{PhysRevLett.107.133602}%
  \BibitemOpen
  \bibfield  {author} {\bibinfo {author} {\bibfnamefont {A.~V.}\ \bibnamefont
  {Gorshkov}}, \bibinfo {author} {\bibfnamefont {J.}~\bibnamefont {Otterbach}},
  \bibinfo {author} {\bibfnamefont {M.}~\bibnamefont {Fleischhauer}}, \bibinfo
  {author} {\bibfnamefont {T.}~\bibnamefont {Pohl}}, \ and\ \bibinfo {author}
  {\bibfnamefont {M.~D.}\ \bibnamefont {Lukin}},\ }\href {\doibase
  10.1103/PhysRevLett.107.133602} {\bibfield  {journal} {\bibinfo  {journal}
  {Phys. Rev. Lett.}\ }\textbf {\bibinfo {volume} {107}},\ \bibinfo {pages}
  {133602} (\bibinfo {year} {2011})}\BibitemShut {NoStop}%
\bibitem [{\citenamefont {Wade}\ \emph {et~al.}(2016)\citenamefont {Wade},
  \citenamefont {Mattioli},\ and\ \citenamefont
  {M\o{}lmer}}]{PhysRevA.94.053830}%
  \BibitemOpen
  \bibfield  {author} {\bibinfo {author} {\bibfnamefont {A.~C.~J.}\
  \bibnamefont {Wade}}, \bibinfo {author} {\bibfnamefont {M.}~\bibnamefont
  {Mattioli}}, \ and\ \bibinfo {author} {\bibfnamefont {K.}~\bibnamefont
  {M\o{}lmer}},\ }\href {\doibase 10.1103/PhysRevA.94.053830} {\bibfield
  {journal} {\bibinfo  {journal} {Phys. Rev. A}\ }\textbf {\bibinfo {volume}
  {94}},\ \bibinfo {pages} {053830} (\bibinfo {year} {2016})}\BibitemShut
  {NoStop}%
\bibitem [{\citenamefont {Saffman}\ \emph {et~al.}(2010)\citenamefont
  {Saffman}, \citenamefont {Walker},\ and\ \citenamefont
  {M\o{}lmer}}]{RevModPhys.82.2313}%
  \BibitemOpen
  \bibfield  {author} {\bibinfo {author} {\bibfnamefont {M.}~\bibnamefont
  {Saffman}}, \bibinfo {author} {\bibfnamefont {T.~G.}\ \bibnamefont {Walker}},
  \ and\ \bibinfo {author} {\bibfnamefont {K.}~\bibnamefont {M\o{}lmer}},\
  }\href {\doibase 10.1103/RevModPhys.82.2313} {\bibfield  {journal} {\bibinfo
  {journal} {Rev. Mod. Phys.}\ }\textbf {\bibinfo {volume} {82}},\ \bibinfo
  {pages} {2313} (\bibinfo {year} {2010})}\BibitemShut {NoStop}%
\bibitem [{\citenamefont {Li}\ \emph {et~al.}(2008)\citenamefont {Li},
  \citenamefont {Yang}, \citenamefont {Cao}, \citenamefont {Zhang},
  \citenamefont {Xie},\ and\ \citenamefont {Wang}}]{PhysRevLett.101.073602}%
  \BibitemOpen
  \bibfield  {author} {\bibinfo {author} {\bibfnamefont {S.}~\bibnamefont
  {Li}}, \bibinfo {author} {\bibfnamefont {X.}~\bibnamefont {Yang}}, \bibinfo
  {author} {\bibfnamefont {X.}~\bibnamefont {Cao}}, \bibinfo {author}
  {\bibfnamefont {C.}~\bibnamefont {Zhang}}, \bibinfo {author} {\bibfnamefont
  {C.}~\bibnamefont {Xie}}, \ and\ \bibinfo {author} {\bibfnamefont
  {H.}~\bibnamefont {Wang}},\ }\href {\doibase 10.1103/PhysRevLett.101.073602}
  {\bibfield  {journal} {\bibinfo  {journal} {Phys. Rev. Lett.}\ }\textbf
  {\bibinfo {volume} {101}},\ \bibinfo {pages} {073602} (\bibinfo {year}
  {2008})}\BibitemShut {NoStop}%
\bibitem [{\citenamefont {Shiau}\ \emph {et~al.}(2011)\citenamefont {Shiau},
  \citenamefont {Wu}, \citenamefont {Lin},\ and\ \citenamefont
  {Chen}}]{PhysRevLett.106.193006}%
  \BibitemOpen
  \bibfield  {author} {\bibinfo {author} {\bibfnamefont {B.-W.}\ \bibnamefont
  {Shiau}}, \bibinfo {author} {\bibfnamefont {M.-C.}\ \bibnamefont {Wu}},
  \bibinfo {author} {\bibfnamefont {C.-C.}\ \bibnamefont {Lin}}, \ and\
  \bibinfo {author} {\bibfnamefont {Y.-C.}\ \bibnamefont {Chen}},\ }\href
  {\doibase 10.1103/PhysRevLett.106.193006} {\bibfield  {journal} {\bibinfo
  {journal} {Phys. Rev. Lett.}\ }\textbf {\bibinfo {volume} {106}},\ \bibinfo
  {pages} {193006} (\bibinfo {year} {2011})}\BibitemShut {NoStop}%
\bibitem [{\citenamefont {Chen}\ \emph {et~al.}(2012)\citenamefont {Chen},
  \citenamefont {Lee}, \citenamefont {Hung}, \citenamefont {Chen},
  \citenamefont {Chen},\ and\ \citenamefont {Yu}}]{PhysRevLett.108.173603}%
  \BibitemOpen
  \bibfield  {author} {\bibinfo {author} {\bibfnamefont {Y.-H.}\ \bibnamefont
  {Chen}}, \bibinfo {author} {\bibfnamefont {M.-J.}\ \bibnamefont {Lee}},
  \bibinfo {author} {\bibfnamefont {W.}~\bibnamefont {Hung}}, \bibinfo {author}
  {\bibfnamefont {Y.-C.}\ \bibnamefont {Chen}}, \bibinfo {author}
  {\bibfnamefont {Y.-F.}\ \bibnamefont {Chen}}, \ and\ \bibinfo {author}
  {\bibfnamefont {I.~A.}\ \bibnamefont {Yu}},\ }\href {\doibase
  10.1103/PhysRevLett.108.173603} {\bibfield  {journal} {\bibinfo  {journal}
  {Phys. Rev. Lett.}\ }\textbf {\bibinfo {volume} {108}},\ \bibinfo {pages}
  {173603} (\bibinfo {year} {2012})}\BibitemShut {NoStop}%
\bibitem [{\citenamefont {Artoni}\ and\ \citenamefont
  {Zavatta}(2015)}]{PhysRevLett.115.113005}%
  \BibitemOpen
  \bibfield  {author} {\bibinfo {author} {\bibfnamefont {M.}~\bibnamefont
  {Artoni}}\ and\ \bibinfo {author} {\bibfnamefont {A.}~\bibnamefont
  {Zavatta}},\ }\href {\doibase 10.1103/PhysRevLett.115.113005} {\bibfield
  {journal} {\bibinfo  {journal} {Phys. Rev. Lett.}\ }\textbf {\bibinfo
  {volume} {115}},\ \bibinfo {pages} {113005} (\bibinfo {year}
  {2015})}\BibitemShut {NoStop}%
\bibitem [{\citenamefont {Liu}\ \emph {et~al.}(2016)\citenamefont {Liu},
  \citenamefont {Chen}, \citenamefont {Chen}, \citenamefont {Lo}, \citenamefont
  {Tsai}, \citenamefont {Yu}, \citenamefont {Chen},\ and\ \citenamefont
  {Chen}}]{PhysRevLett.117.203601}%
  \BibitemOpen
  \bibfield  {author} {\bibinfo {author} {\bibfnamefont {Z.-Y.}\ \bibnamefont
  {Liu}}, \bibinfo {author} {\bibfnamefont {Y.-H.}\ \bibnamefont {Chen}},
  \bibinfo {author} {\bibfnamefont {Y.-C.}\ \bibnamefont {Chen}}, \bibinfo
  {author} {\bibfnamefont {H.-Y.}\ \bibnamefont {Lo}}, \bibinfo {author}
  {\bibfnamefont {P.-J.}\ \bibnamefont {Tsai}}, \bibinfo {author}
  {\bibfnamefont {I.~A.}\ \bibnamefont {Yu}}, \bibinfo {author} {\bibfnamefont
  {Y.-C.}\ \bibnamefont {Chen}}, \ and\ \bibinfo {author} {\bibfnamefont
  {Y.-F.}\ \bibnamefont {Chen}},\ }\href {\doibase
  10.1103/PhysRevLett.117.203601} {\bibfield  {journal} {\bibinfo  {journal}
  {Phys. Rev. Lett.}\ }\textbf {\bibinfo {volume} {117}},\ \bibinfo {pages}
  {203601} (\bibinfo {year} {2016})}\BibitemShut {NoStop}%
\bibitem [{\citenamefont {Hammerer}\ \emph {et~al.}(2010)\citenamefont
  {Hammerer}, \citenamefont {S\o{}rensen},\ and\ \citenamefont
  {Polzik}}]{RevModPhys.82.1041}%
  \BibitemOpen
  \bibfield  {author} {\bibinfo {author} {\bibfnamefont {K.}~\bibnamefont
  {Hammerer}}, \bibinfo {author} {\bibfnamefont {A.~S.}\ \bibnamefont
  {S\o{}rensen}}, \ and\ \bibinfo {author} {\bibfnamefont {E.~S.}\ \bibnamefont
  {Polzik}},\ }\href {\doibase 10.1103/RevModPhys.82.1041} {\bibfield
  {journal} {\bibinfo  {journal} {Rev. Mod. Phys.}\ }\textbf {\bibinfo {volume}
  {82}},\ \bibinfo {pages} {1041} (\bibinfo {year} {2010})}\BibitemShut
  {NoStop}%
\bibitem [{Sup()}]{SuppInfo}%
  \BibitemOpen
  \href@noop {} {\bibinfo  {journal} {See Supplemental Material at URL for
  details of derivations and more examples from numerical simulations}\
  }\BibitemShut {NoStop}%
\bibitem [{\citenamefont {Fleischhauer}\ \emph {et~al.}(2005)\citenamefont
  {Fleischhauer}, \citenamefont {Imamoglu},\ and\ \citenamefont
  {Marangos}}]{RevModPhys.77.633}%
  \BibitemOpen
\bibfield  {journal} {  }\bibfield  {author} {\bibinfo {author} {\bibfnamefont
  {M.}~\bibnamefont {Fleischhauer}}, \bibinfo {author} {\bibfnamefont
  {A.}~\bibnamefont {Imamoglu}}, \ and\ \bibinfo {author} {\bibfnamefont
  {J.~P.}\ \bibnamefont {Marangos}},\ }\href {\doibase
  10.1103/RevModPhys.77.633} {\bibfield  {journal} {\bibinfo  {journal} {Rev.
  Mod. Phys.}\ }\textbf {\bibinfo {volume} {77}},\ \bibinfo {pages} {633}
  (\bibinfo {year} {2005})}\BibitemShut {NoStop}%
\bibitem [{\citenamefont {Deng}\ and\ \citenamefont
  {Payne}(2005)}]{PhysRevA.71.011803}%
  \BibitemOpen
  \bibfield  {author} {\bibinfo {author} {\bibfnamefont {L.}~\bibnamefont
  {Deng}}\ and\ \bibinfo {author} {\bibfnamefont {M.~G.}\ \bibnamefont
  {Payne}},\ }\href {\doibase 10.1103/PhysRevA.71.011803} {\bibfield  {journal}
  {\bibinfo  {journal} {Phys. Rev. A}\ }\textbf {\bibinfo {volume} {71}},\
  \bibinfo {pages} {011803} (\bibinfo {year} {2005})}\BibitemShut {NoStop}%
\bibitem [{\citenamefont {Sun}\ and\ \citenamefont
  {Metcalf}(2014)}]{PhysRevA.90.033408}%
  \BibitemOpen
  \bibfield  {author} {\bibinfo {author} {\bibfnamefont {Y.}~\bibnamefont
  {Sun}}\ and\ \bibinfo {author} {\bibfnamefont {H.}~\bibnamefont {Metcalf}},\
  }\href {\doibase 10.1103/PhysRevA.90.033408} {\bibfield  {journal} {\bibinfo
  {journal} {Phys. Rev. A}\ }\textbf {\bibinfo {volume} {90}},\ \bibinfo
  {pages} {033408} (\bibinfo {year} {2014})}\BibitemShut {NoStop}%
\end{thebibliography}%
\end{document}